\documentclass[12pt]{article}

\def\ms{m_{S}}
\def\msoft{m_{\rm soft}}
\def\qs{Q_S'}
\def\qh{Q_H'}
\def\mew{m_{\rm EW}}
\def\mzp{M_{Z'}}
\def\uop{U(1)'}
\def\zp{Z'}
\def\MPL #1 #2 #3 {Mod.~Phys.~Lett.~{\bf#1},\  #2 (#3)}
\def\NPB #1 #2 #3 {Nucl.~Phys.~{\bf#1},\  #2 (#3)}
\def\PLB #1 #2 #3 {Phys.~Lett.~{\bf#1},\  #2 (#3)}
\def\PR #1 #2 #3 {Phys.~Rep.~{\bf#1},\ #2 (#3)}
\def\PRD #1 #2 #3 {Phys.~Rev.~{\bf#1},\  #2 (#3)}
\def\PRL #1 #2 #3 {Phys.~Rev.~Lett.~{\bf#1},\  #2 (#3)}
\def\RMP #1 #2 #3 {Rev.~Mod.~Phys.~{\bf#1},\  #2 (#3)}
\def\ZP #1 #2 #3 {Z.~Phys.~{\bf#1},\  #2 (#3)}
\def\IJMP #1 #2 #3 {Int.~J.~Mod.~Phys.~{\bf#1},\  #2 (#3)}
\def\PROG #1 #2 #3 {Prog.~Theo.~Phys.~{\bf#1},\  #2 (#3)}

\def\gsim{\mathrel{\raise.3ex\hbox{$>$\kern-.75em\lower1ex\hbox{$\sim$}}}}
\begin{document}
\begin{titlepage}
\hbox{\hfil\hskip 4in FERMILAB-CONF-96/344-T}
\hbox{\hfil\hskip 4in October 1, 1996}
\vskip .5in
\centerline{\LARGE $Z'$ Bosons and Supersymmetry\footnote{
To appear in the Proceedings of the 1996 DPF/DPB
Summer Study on New Directions for High Energy Physics
(Snowmass 96), Snowmass CO, June 25 - July 12, 1996.}}
\vskip .5in
\centerline{\large
Joseph D. Lykken\footnote{Research supported by the U.S. Department of
Energy under contract DE-AC03-76SF00515.}}
\vskip .3in
\centerline{
Theoretical Physics Dept., MS106}
\centerline{
Fermi National Accelerator Laboratory}
\centerline{
P.O. Box 500, Batavia, IL 60510}
\vskip .5in

\begin{abstract}
A broad class of supersymmetric extensions of the
standard model predict a $Z'$ vector boson whose
mass is {\it naturally} in the range
$250$ GeV$<$$\mzp$$<$$2$ TeV. To avoid
unacceptably large mixing with the $Z$, one
requires either a discrete tuning of the $U(1)'$
charges or a leptophobic $Z'$. Both cases are likely
to arise as the low energy limits of heterotic
string compactifications, but a survey of existing
realistic string models provides no acceptable
examples. A broken $U(1)'$ leads to additional
D-term contributions to squark, slepton, and Higgs
masses, which depend on the $U(1)'$ charge assignments
and the $Z'$ mass. The Tevatron and future colliders
can discover or decisively rule out this class of models.
\bigskip\bigskip
\end{abstract}
\end{titlepage}

\section{Introduction}
The minimal extension of the standard model (SM) gauge group
is to append an abelian factor $\uop$.
If at least some of the SM particles have nonzero
$\uop$ charges, the $\uop$ gauge
symmetry must be spontaneously or dynamically broken at
some scale greater than the weak scale, leading to a
massive $\zp$ vector boson which decays into SM particles
and mixes with the SM $Z$ boson.
The existence of such broken $\uop$ gauge symmetries is
a natural prediction of grand unification schemes
like $SO(10)$ and $E_6$, as well as superstring theory.

It is important to distinguish between models in which
the mechanism of $\uop$ breaking is linked to the
mechanism of electroweak symmetry breaking (EWSB),
and models in which the mechanism of $\uop$ breaking
is independent of, or merely parallels, that of EWSB.
An example of the first case is a technicolor theory with
a $U(1)$ factor in the technicolor gauge group.
Here obviously the $\zp$ mass is tied to the technicolor
scale and thus to the electroweak scale. Although a
technicolor $\uop$ has no tree level couplings to
SM particles, it is in principle observable due to
loop or strong interaction effects \cite{moreholdom}.
An example of
the second case is a grand unified (GUT) theory in
which $\uop$ breaking is triggered by the
renormalization group (RG) evolution of (exotic)
Yukawa couplings driving some (exotic) scalar mass
squared negative. Here the mechanism of $\uop$
breaking parallels a possible mechanism of EWSB,
but due to the logarithmic nature of the RG evolution,
the $\zp$ mass is highly unlikely to be within 1 or 2
orders of magnitude of the $Z$ mass without considerable
tuning of the model. Indeed generically $\mzp$ lies in the
range $10^8$ to $10^{16}$ GeV for GUT models.

In this regard, it is interesting to examine the
status of $\zp$ bosons in models of
{\it weak-scale supersymmetry}. This is the
class of models which embed the supersymmetrized
standard model and tie EWSB to supersymmetry breaking
(usually via the dynamics of some new ``messenger'' fields).
This includes the minimal supergravity models \cite{mles}
(also called
CSSM, MLES, and occasionally (improperly) MSSM), as well
as the gauge-mediated low energy breaking
models (GMLESB) \cite{gmlesb}.
In weak-scale supersymmetry the effective supersymmetry
(SUSY) breaking scale in the ``visible'' sector
is the same as the weak scale, $\mew = 246$ GeV.
Typically the
up and down-type Higgs, the squarks, sleptons, and gauginos
all get soft SUSY breaking mass terms which
are of order $\mew ^2$.

In workable models \cite{nilles}
EWSB occurs radiatively, i.e., the up-type Higgs mass
squared is driven negative in the RG evolution, due to
the large top quark Yukawa.
In weak scale supersymmetry models with a $\zp$
it is typical that the $\uop$ breaking is also
radiative, i.e. it is triggered either by a scalar
mass squared going negative or condensate formation,
both mechanisms being driven by RG evolution.
Cvetic and Langacker \cite{cvetic} have identified the subset of
models for which the $\zp$ is naturally light --
within an order of magnitude of the weak scale.
This subset of weak scale supersymmetry models is
the subject of this report. One should note,
however, that in SUSY models even a very heavy $\zp$
does not always completely decouple from collider physics,
because the $\uop$ breaking induces
D-term contributions to scalar masses.

Direct production of a $\zp$, followed by decay to electrons,
muons, or jets, will be observable at the LHC for
$\mzp$ $\leq$ about 5 TeV, assuming roughly SM strength
couplings \cite{search,subgroup}.
The most stringent current bounds come from
the CDF (preliminary) analysis of
$\zp \rightarrow ee$, $\mu\mu$ in 110pb$^{-1}$ of
Tevatron collider data \cite{newcdf}.
Assuming SM couplings to leptons
produces the bound $\mzp > 690$ GeV. For a
``leptophobic'' $\zp$, a mass bound is obtained from
the dijet channel, i.e. searching for resonant structure
in the dijet invariant mass spectrum,
assuming that the $\zp$ width is not too large.
The current limit from UA2 for SM couplings is \cite{ua}
$\mzp > 237$ GeV; a CDF dijet analysis \cite{cdf} excludes
$\zp$ bosons over a wide mass range up to about 1 TeV,
but only for couplings which are considerably more than
SM strength\footnote{The CDF data do not extend the UA2
bound for SM strength couplings essentially because of
extensive prescaling for dijet masses below 353 GeV.}.
Besides direct
production, $Z$-$\zp$ mixing implies effects on the
oblique parameters S,T, and U, as well as other
precision electroweak observables \cite{moreholdom}.
The current LEP data provides strong constraints on
these effects; depending on the $\zp$ couplings
to SM fermions, one can rule out $\mzp$ as large
as 400 GeV, and $Z$-$\zp$ mixing angles greater than
a few times $10^{-3}$ \cite{cvetic}.

\section{Supersymmetry and $\uop$ Breaking}

The $\uop$ gauge symmetry is broken by nonzero
vevs of the scalar components of some chiral
superfields which have nonzero $\uop$ charge.
These chiral superfields may represent either
fundamental particles or composites (or both).
Since we are only interested in $\uop$ breaking
in the visible sector, this vacuum state must be
continuously connected to a SUSY-preserving vacuum,
reached in the limit that the soft SUSY breaking
terms are turned off. Thus we may always assume
that the $\uop$ breaking vacuum lies in some
F and D flat direction, modulo
corrections of order the soft breaking scale, $\msoft$.

Since the scalars which get vevs carry (at least)
a nonzero $\uop$ charge, D flatness implies that
there are at least two fields which get vevs,
modulo \footnote{Note that
``modulo'' includes the possibility that only one
scalar gets a $\uop$ breaking vev, as long as this vev
is no larger than order $\msoft$, and vanishes when
the soft breaking is turned off.}
corrections of order $\msoft$. At least one
of these fields must be a SM singlet, since the
scenario in which
$\uop$ is broken just by the supersymmetric SM
Higgs $H_U$ and $H_D$ is phenomenologically untenable.
There are thus two classes of models to consider:

\begin{itemize}
\item Models with two or more SM singlets getting
$\uop$ breaking vevs.
\item Models with a single SM singlet, $S$, getting a
$\uop$ breaking vev. In these models D flatness
requires one or both of the SM Higgs $H_U$ and
$H_D$ has a nonzero $\uop$ charge.
\end{itemize}

In the first case, D flatness imposes a relation
between the SM singlet vevs, but does not fix
the overall scale. Instead, $\mzp$ is determined
by the RG evolution which (by assumption) drives
the singlet mass squareds negative, together with the
corrections to D-flatness of order $\msoft$.

In the second class of models, D flatness implies
that the vev of $S$ is related to the Higgs vevs
(with coefficients that are just the $\uop$ charges),
modulo corrections of order $\msoft$. Let us suppose
that the soft breaking mass $\ms$ associated with $S$ is
of order the weak scale. This will in fact be true
automatically in most models, since the SM soft breaking
terms are of order $\mew$. For this subset of weak
scale supersymmetry models it is clear that the vev
of $S$ is of order $\mew$, since
\begin{equation}
\ms \simeq \msoft \simeq \mew .
\end{equation}
The resulting value of
$\mzp$ is a function of gauge couplings and the
parameters
\begin{equation}
\ms, \mew, \qs, \qh,
\end{equation}
where the latter two are the $\uop$ charges of
$S$ and the Higgs doublet(s). $\mzp$ is naturally
within an order of magnitude of $\mew$ in this class
of models, which I will refer to as
``Cvetic-Langacker'' models.

Since we have assumed that $S$ gets a vev,
the RG evolution must drive $\ms ^2$ negative at some
scale $\Lambda$. This does not imply, however, that
$\Lambda$ is anywhere near $\mew$; it may be many
orders of magnitude larger. Nevertheless supersymmetry
ensures that the vev of $S$ is of order $\mew$, not
of order $\Lambda$.

As emphasized by Cvetic and Langacker \cite{cvetic},
the main
phenomenological deficiency of this class of models
is that the $Z$-$\zp$ mixing angle is not
sufficiently suppressed for $\mzp \simeq 1$ TeV.
This problem is avoided in two cases:
\begin{itemize}
\item The $\zp$ is
leptophobic, i.e. the $Q'$ charges of the SM
leptons vanish.
\item $\qs$ and $\qh$ have the same relative sign
and take values in a certain narrow range.
\end{itemize}
Either case can be considered a discrete tuning
of the $\uop$ charges. Thus although the above
discussion relied only on general properties of
weak scale supersymmetry, to get a completely
natural scenario we must embed these models in
a larger framework like GUT's or superstrings.

\section{Superstring Models}
For purposes of studying phenomenological prospects
at future colliders, it would be useful to have
one or more ``benchmark'' models with naturally
light $\zp$'s. Because of the $Z$-$\zp$ mixing
problem, an obvious place to look is among
the known four-dimensional
$N$$=$$1$ spacetime supersymmetric solutions to the
weakly coupled heterotic superstring.

Roughly two dozen heterotic string vacua have been
constructed which are realistic in the sense that
they embed the SM gauge group along with three
generations of SM fermions and some number of vectorlike
exotics. These string models often contain one or
more $\uop$'s which remain unbroken at the string scale,
and a number of SM singlet fields which have nonzero
charges under $\uop$. The $\uop$ charges of all particles,
including SM particles, are fixed.
Each string model is actually
a continuous multiparameter family of string vacua,
depending on the values of moduli vevs which are not determined
in string perturbation theory.
Further, although these string models contain
roughly the right ingredients for hidden sector
dynamical SUSY breaking, no one has as yet
performed a detailed analysis of SUSY breaking
and the resulting soft breaking terms for a
complete realistic string model.
Nevertheless given the fixed $\uop$ charges and
using various string consistency conditions, one
can determine whether the Cvetic-Langacker scenario
is at least possible in a given string model.

Cvetic and Langacker surveyed some existing string
models for cases which employ the second,
leptophilic, solution of the mixing problem.
They found no acceptable candidates. This is
not surprising given that only a handful of models
were looked at, and that this solution requires a
a tuning of $\qs$ and $\qh$.

I have performed a similar survey of half a dozen
string models, this time looking for
Cvetic-Langacker in the leptophobic mode.
The results are shown in Table 1.

\begin{table}
\begin{center}
\begin{tabular}{c|cc}
\hline
&\vrule height4pt width0pt depth0pt&\cr
Model & Leptophobic $\uop$? & $\qh$ $\neq$ 0? \cr
&\vrule height4pt width0pt depth0pt&\cr
\hline
&\vrule height2pt width0pt depth0pt&\cr
Faraggi I\ \cite{farI}&no&-- \cr
&\vrule height2pt width0pt depth0pt&\cr
Faraggi II\ \cite{farII}&yes&no \cr
&\vrule height2pt width0pt depth0pt&\cr
Faraggi et al\ \cite{fny}&no&-- \cr
&\vrule height2pt width0pt depth0pt&\cr
Chaudhuri et al\ \cite{chl}&yes&no \cr
&\vrule height2pt width0pt depth0pt&\cr
Hockney-Lykken\ \cite{hl}&yes&no \cr
&\vrule height2pt width0pt depth0pt&\cr
Flipped SU(5)\ \cite{flip}&yes&yes\cr
&\vrule height2pt width0pt depth0pt&\cr
\hline
\end{tabular}
\caption{Partial survey of string models
for leptophobic Cvetic-Langacker candidates.}
\label{one}
\end{center}
\end{table}

This sampling of models is sufficient to draw two
major conclusions:
\begin{itemize}
\item It is not difficult to construct realistic
string models with a leptophobic $\uop$ unbroken
at the string scale. This observation has already
been made in the literature \cite{farlep,loplep}.
\item The leptophobic string models are unlikely
to be Cvetic-Langacker models, because typically
the Higgs doublets are uncharged under the
leptophobic $\uop$. The exception to this rule
in Table 1 is the flipped $SU(5)$ model. The
reason for this is rather elementary: the existing
realistic string models have an underlying $E_6$,
$SO(10)$, or $SU(5)$ gauge structure built in,
broken in a stringy way by Wilson lines. In order
to have nonvanishing Yukawa couplings, the Higgs
doublets typically have nonvanishing charge only
under these $E_6$ based $U(1)$s (in the simplest
models, the Higgs arise from the untwisted sector).
It is well known that within $E_6$ the only
possibility for symmetry-based leptophobia is
flipped $SU(5)$\footnote{If there is $Z$-$\zp$
or photon-$\zp$ mixing in the kinetic terms of
the effective field theory, then leptophobia is
still possible within $E_6$ \cite{holdom,babu,rosner}.
This involves subtle issues regarding the effective
gauge kinetic function which are beyond the scope of
this report.}.
\end{itemize}

Unfortunately flipped $SU(5)$ does not provide an
acceptable example of a Cvetic-Langacker model either. Having
fixed the particle identification in the usual way
so as to generate a large top quark Yukawa, one finds
that the first and second generation quarks have
different charges under the leptophobic $\uop$.
This would lead to flavor-changing neutral
currents\footnote{This is not necessarily a
disaster if the $\zp$ charge eigenstate down-type quarks
are also mass eigenstates: see the revised version
of \cite{loplep}.}.

\section{D-Term Contributions to Scalar Masses}

As mentioned in the introduction, even a very heavy
$\zp$ boson does not completely decouple from
collider physics in a supersymmetric theory \cite{drees}.
This is because a D-term contribution is generated
to the scalar potential:
\begin{equation}
V_{D} = {g^2\over 2}\left[ \sum_i Q_i \vert \phi _i
\vert ^2 \right]^2,
\end{equation}
\noindent where $g$ is the $\uop$ gauge coupling,
$\phi _i$ are all the scalar fields, and $Q_i$ are
their $\uop$ charges.
Since some of the scalars must get vevs to break the
$\uop$, every scalar mass squared receives a
contribution of the form \cite{susyrep}
\begin{equation}
\Delta m_i^2 = Q_i \Lambda ^2,
\end{equation}
\noindent where $\Lambda$ is an overall
scale\footnote{It may appear that the D-terms also
imply new contributions to the quartic scalar couplings;
that this is false can be seen by explicitly integrating
out the fields whose vevs break $\uop$.}.

In the Cvetic-Langacker scenario these D-term
contributions are given by the approximate expression:
\begin{equation}
\Delta m_i^2 = Q_i {\mzp ^2\over 2Q_S}.
\end{equation}

These splittings are roughly of order $\pm$(250 GeV)$^2$.
Thus if supersymmetry is observed in future collider
experiments, a Cvetic-Langacker $\zp$ implies large
deviations in the sparticle mass spectrum from the
patterns characteristic either of minimal supergravity
or of GMLESB.

\section{Conclusion}
Within the context of weak scale supersymmetry there
is a broad class of models which predict a $\zp$
boson whose mass cannot be much more than 1 TeV.
This prediction is natural given the
usual assumptions of weak scale supersymmetry.
To achieve in addition a large natural suppression
of $Z$-$\zp$ mixing, these models should be embedded
in some larger framework such as superstrings.
There are some obstacles to providing acceptable
superstring models, but they do not seem
insurmountable. A ``benchmark'' model, i.e. a
specific realization of the Cvetic-Langacker scenario
with fixed $\uop$ charge assignments, has at most
one new free parameter compared to minimal supergravity
or whatever version of weak scale SUSY it is embedded in.
However such a model has many new observables:
the $\zp$ mass, width, and branching fractions to
SM and sparticle decay modes, as
well as the observable effects of $Z$-$\zp$ mixing.
These observables can be used to provide
overconstrained predictions of the D-term
contributions to scalar masses.

If these models are realized in the leptophobic mode,
the $\zp$ resonance must still show up in the dijet
spectrum at either the Tevatron or the LHC. A SUSY
$\zp$ discovery would be strong motivation towards
running a high luminosity NLC at the $Z$ pole, and
towards building a muon collider which could operate
at the $\zp$ pole.



\begin{thebibliography}{99}

\bibitem{moreholdom}
B. Holdom, \PLB 259 329 1991 .

\bibitem{mles}
A. Chamseddine, R. Arnowitt, and P. Nath,
\PRL 49 970 1982 ;
R. Barbieri, S. Ferrara, and C. Savoy,
\PLB 119 343 1982 ;
L. Hall, J. Lykken, and S. Weinberg,
\PRD 27 2359 1983 .

\bibitem{gmlesb}
M. Dine and W. Fischler, \PLB 110 227 1982 ;
C. Nappi and B. Ovrut, \PLB 113 1751 1982 ;
K. Inoue et al, \PROG 67 1889 1982 ;
J. Ellis, L. Ibanez, and G. Ross, \PLB 113 283 1982 .

\bibitem{nilles}
H.P. Nilles, \PR 110 1 1984 .

\bibitem{cvetic}
M. Cvetic and P. Langacker, \PRD 54 3570 1996 ;
\MPL A11 1247 1996 .

\bibitem{search}
``Searches for new gauge bosons at future colliders'',
T.G. Rizzo, SLAC-PUB-7279, hep-ph/9609248;
``Determination of the $Z'$ mass and couplings
below threshold at the NLC'', SLAC-PUB-7250, hep-ph/9608274.

\bibitem{subgroup}
Report of the New Gauge Bosons subgroup of the New
Phenomena Working Group, Proceedings of the
1996 DPF/DPB Summer Study on New Directions for High
Energy Physics (Snowmass 96), Snowmass CO, June 25 - July 12, 1996.

\bibitem{newcdf}
M. Pillai et al, the CDF collaboration, ``A search for
neutral heavy vector gauge bosons in
$\bar{p}p$ collisions at $\sqrt{s} = 1.8$ TeV'',
FERMILAB-CONF-96/279-E, Proceedings of the
1996 Divisional Meeting of the DPF, American Physical
Society, Minneapolis, MN, Aug. 10-15 1996.

\bibitem{ua}
J. Alitti et al, the UA2 collaboration,
\NPB 400 3 1993 .

\bibitem{cdf}
F. Abe et al, the CDF collaboration,
\PRL 74 3539 1995 .

\bibitem{farlep}
A. Faraggi and M. Masip, ``Leptophobic $\zp$ from
superstring derived models'', hep-ph/9604302.

\bibitem{loplep}
J. Lopez and D. Nanopoulos, ``Leptophobic $\zp$
in stringy flipped $SU(5)$'', hep-ph/9605359.

\bibitem{farI}
A. Faraggi, \PLB 278 131 1992 .

\bibitem{farII}
A. Faraggi, \PLB 339 223 1994 .

\bibitem{fny}
A. Faraggi, D. Nanopoulos, and K. Yuan, \NPB 335 347 1990 .

\bibitem{chl}
S. Chaudhuri, G. Hockney, and J. Lykken,
\NPB 469 357 1996 .

\bibitem{hl}
J. Lykken, ``String model building in the Age of D-branes'',
talk at the 4th International Conference on Supersymmetries
in Physics (SUSY 96), College Park,MD,29 May - 1 Jun 1996,
hep-th/9607144.

\bibitem{flip}
J. Lopez, D. Nanopoulos, and K. Yuan, \NPB 399 654 1993 ;
an earlier version is I. Antoniadis, J. Ellis, J. Hagelin,
and D. Nanopoulos, \PLB 231 65 1989 .

\bibitem{holdom}
B. Holdom, \PLB 166 196 1986 ;
F. del Aguila, G. Coughlan, and M. Quiros,
\NPB 307 633 1988 .

\bibitem{babu}
K. Babu, C. Kolda, and J. March-Russell,
\PRD 54 4635 1996 .

\bibitem{rosner}
J. Rosner, ``Prominent decay modes of a leptophobic
$\zp$'', hep-ph/9607207.

\bibitem{drees}
M. Drees, \PLB 181 279 1986 ; H.-C. Cheng and L. Hall,
\PRD 51 1337 1995 ; C. Kolda and S. Martin,
\PRD 53 3871 1996 .

\bibitem{susyrep}
J. Amundson et al, Report of the Supersymmetry Theory
Subgroup, Proceedings of the
1996 DPF/DPB Summer Study on New Directions for High
Energy Physics (Snowmass 96), Snowmass CO, June 25 - July 12, 1996;
hep-ph/9609374.

\end{thebibliography}
\end{document}